\documentclass[10pt,twocolumn]{article}

\usepackage[utf8]{inputenc}
\usepackage{amsmath}
\usepackage{amssymb}
\usepackage{amsthm}
\usepackage{titlesec}
\usepackage{natbib}
\usepackage{booktabs}
\usepackage{graphicx}
\usepackage[colorlinks=false,hidelinks]{hyperref}
\usepackage[format=plain,labelfont=it]{caption}
\usepackage[left=1.5cm,right=1.5cm,top=2cm,bottom=2cm]{geometry}

\newtheorem{defn}{Definition}
\newtheorem{rem}{Remark}
\newtheorem{prop}{Proposition}

\titleformat{\section}{\centering\normalfont\scshape}{\arabic{section}.}{5pt}{}
\titleformat{\subsection}{\normalfont\it}{\arabic{section}.\arabic{subsection}}{5pt}{}
\titleformat{\subsubsection}{\normalfont\it}{\arabic{section}.\arabic{subsection}.\arabic{subsubsection}}{5pt}{}

\newcommand\infoFootnote[1]{%
  \begingroup
  \renewcommand\thefootnote{}\footnote{#1}%
  \addtocounter{footnote}{-1}%
  \endgroup}

\newcommand{\F}{\mathbb{F}}
\newcommand{\N}{\mathbb{N}} 
\newcommand{\R}{\mathbb{R}}

\newcommand{\rb}{\boldsymbol{r}}
\renewcommand{\sb}{\boldsymbol{s}}
\newcommand{\ub}{\boldsymbol{u}}
\newcommand{\xb}{\boldsymbol{x}}
\newcommand{\yb}{\boldsymbol{y}}
\newcommand{\zb}{\boldsymbol{z}}

\newcommand{\alphab}{\boldsymbol{\alpha}}

\newcommand{\zerob}{\boldsymbol{0}}
\newcommand{\oneb}{\boldsymbol{1}}

\newcommand{\Ab}{\boldsymbol{A}}
\newcommand{\Bb}{\boldsymbol{B}}
\newcommand{\Cb}{\boldsymbol{C}}
\newcommand{\Hb}{\boldsymbol{H}}
\newcommand{\Ib}{\boldsymbol{I}}
\newcommand{\Nb}{\boldsymbol{N}}
\newcommand{\Rb}{\boldsymbol{R}}
\newcommand{\Sb}{\boldsymbol{S}}
\newcommand{\Vb}{\boldsymbol{V}}
\newcommand{\Wb}{\boldsymbol{W}}

\newcommand{\Upsilonb}{\boldsymbol{\Upsilon}}

\newcommand{\Cc}{\mathcal{C}}
\newcommand{\Mc}{\mathcal{M}}
\newcommand{\Rc}{\mathcal{R}}
\newcommand{\Uc}{\mathcal{U}}

\newcommand{\vect}[1]{\mathrm{vec}\left(#1\right)}
\newcommand{\floor}[1]{\left\lfloor#1\right\rfloor}

\newcommand{\Dec}{\mathsf{Dec}}
\newcommand{\Enc}{\mathsf{Enc}}
\newcommand{\negl}{\mathsf{negl}}

\title{\vspace{-2mm}\bf Cryptanalysis of Random Affine \\Transformations for Encrypted Control$^\ast$}
\author{Nils Schl\"uter, Philipp Binfet and Moritz Schulze Darup\vspace{2mm}}
\date{}

\begin{document}

\maketitle

\textbf{\textit{Abstract}.} {\bf Cloud-based and distributed computations are of growing interest in modern control systems. However, these technologies require performing computations on not necessarily trustworthy platforms and, thus, put the confidentiality of sensitive control-related data at risk. Encrypted control has dealt with this issue by utilizing modern cryptosystems with homomorphic properties, which allow a secure evaluation at the cost of an increased computation or communication effort (among others). Recently, a cipher based on a random affine transformation gained attention in the encrypted control community. Its appeal stems from the possibility to construct security providing homomorphisms that do not suffer from
the restrictions of ``conventional'' approaches. 
\\
This paper provides a cryptanalysis of random affine transformations in the context of encrypted control. To this end, a deterministic and probabilistic variant of the cipher over real numbers are analyzed in a generalized setup, where we use cryptographic definitions for security and attacker models. It is shown that the deterministic cipher breaks under a known-plaintext attack, and unavoidably leaks information of the closed-loop, which opens another angle of attack. For the probabilistic variant, statistical indistinguishability of ciphertexts can be achieved, which makes successful attacks unlikely. 
We complete our analysis by investigating a floating point realization of the probabilistic random affine transformation cipher, which unfortunately suggests the impracticality of the scheme if a security guarantee is needed.}
\infoFootnote{Nils Schl\"uter, Philipp Binfet and M. Schulze Darup are with the \href{https://rcs.mb.tu-dortmund.de/}{Control and~Cyber-physical Systems Group}, Faculty of Mechanical Engineering, TU Dortmund University, Germany. E-mails: \href{mailto:moritz.schulzedarup@tu-dortmund.de}{\{nils.schlueter, philipp.binfet, moritz.schulzedarup\}@tu-dortmund.de}. \vspace{0.5mm}}
\infoFootnote{N. Schl\"uter and P. Binfet share the first authorship.}
\infoFootnote{\hspace{-1.5mm}$^\ast$This paper is a \textbf{preprint} of a contribution to the 22nd World Congress of the International Federation of Automatic Control 2023.}

\section{Motivation}
The increasing use of reliable wireless communication with low latency for industrial purposes fuels many promising fields in which control and decision-making play a key role, e.g., industry~4.0, smart grids, building automation, robot swarms, and intelligent transportation systems. These cyber-physical systems come along with significant cost reductions, high flexibility, and an improved performance, while one defining feature is that control-related data is exchanged and processed non-locally. This data often contains system information and design specifications, such that the possibility of being hacked or the use of untrustworthy platforms pose threats to the confidentiality.
This is especially worrying in the context of cyber-physical systems, because access to control data enables the construction of effective attacks, such as stealthy attacks, that can have severe consequences.

Confidentiality of control data, e.g., process data and controller parameters, is the major concern in encrypted control (see \citet{SchulzeDarup2021_CSM} for an overview). Here, the enabling technologies are modern cryptosystems with homomorphic properties that allow to evaluate mathematical operations on encrypted data. Consequently, also cloud-based services for different tasks become conceivable despite the data's sensitivity. Many steps towards the practicality of encrypted control have already been made, such as in model predictive control (\citet{SchulzeDarup2018_LCSS,Alexandru2018_CDC,tjell2021secure}), linear dynamic control (\citet{kim2021dynamic,SchlueterDarupECC2021}), or control design (\citet{schluter2022encrypted}).
These solutions build on homomorphic encryption (see \citet{chillotti2020tfhe,Cheon2017_CKKS}) and/or secure multi-party computation and secret sharing (see \citet{cramer2015secure}). Besides that, approaches based on differential privacy (\citet{dwork2014algorithmic}) exist. Of course, the various schemes have different pros and cons and are
suitable for different applications.
A commonality between them is, however, that iterative processes are non-trivial to implement, i.e., homomorphic encryption needs costly bootstrapping, garbled circuits (in secure multi-party computation) cannot be reused, secret sharing needs rescaling protocols, and the noise, added in differential privacy, eats away the accuracy of the result in each iteration. 

Remarkably, this is not the case in random affine transformation (RT) ciphers, also called random matrix encryption or affine masking, which lately gained attention in the encrypted control community. 
This advantage comes from the construction of homomorphisms
over real numbers (instead of, e.g., finite sets). There, the involved randomness provides security, while equivalency of the results is ensured. As a consequence, these schemes provide great flexibility and performance improvements compared to conventional approaches.
As a matter of fact, it seems that an RT has first been suggested for matrix products 
in~(\citet{shan2018practical,lei2014achieving,du2002practical}).
More recently, RTs are first and foremost applied in the context of optimization problems. For application in linear programming, see~\citet{vaidya2009privacy} or~\citet{dreier2011practical}. Linear and nonlinear model predictive control are discussed in~\citet{naseri2022privacy} and~\citet{zhang2021privacy}, respectively. Next, collaborative learning is addressed in~\citet{hayati2022privacy}. The papers by~\citet{weeraddana2013per} and~\citet{sultangazin2020symmetries} put forward a more general analysis. Finally, in~\citet{xu2015secure,wang2011secure}, not only the confidentiality but simultaneously also the integrity of the data is addressed.

Throughout this paper, we will analyze the security of RTs in several scenarios.
Nonetheless, there already exist security certificates for RTs in the literature (see, for instance,~\citet{weeraddana2013per,sultangazin2020symmetries,zhang2021privacy}).
However, the definitions used therein are new or partially adapted from differential privacy, which provides less security (especially when data utility is necessary) in comparison to standard cryptographic definitions. Moreover, many authors consider merely a ciphertext-only attack, which is the weakest attacker model, while control data is often highly correlated and keys tend to be reused.
Motivated by the consequences of potential breaks, we prefer an analysis based on the more strict cryptographic definitions, which has not yet been provided for RTs.
We stress that the purpose of this paper is to inform about the security of RTs (and hopefully provide suitable access to the cryptographic subtleties for control engineers) in several contexts, and, in the spirit of cryptanalysis, to serve as a basis for discussion as well as further improvements.

\textit{Roadmap.}
Section~\ref{sec:setup} mainly introduces some ideas from cryptography and partially adapts them for our purposes. Then, Section~\ref{sec:RT} deals with the RTs as a preparation for the analysis in Sections~\ref{sec:detanalysis}--\ref{sec:floatsec}. Finally, the paper is concluded in Section~\ref{sec:outlook}.

\textit{Notation.}
Frequently we make use of the ``vectorization trick'' $\vect{\Ab\Bb\Cb}=(\Cb^\top \otimes\,\Ab)\vect{\Bb}$, where $\Ab,\Bb$, and $\Cb$ are conformal matrices, $\otimes$ denotes the Kronecker product, and $\vect{\Bb}$ refers to stacking the columns of $\Bb$ in a vector. By $P_d(\cdot)$ and $p_c(\cdot)$ we denote the probability mass function and probability density function of a discrete random variable $d$ and continuous random variable $c$, respectively. We write $P(\cdot)$ and $p(\cdot)$ whenever it is clear from context to which variables these functions apply.

\section{Cryptographic Preliminaries}
\label{sec:setup}
In this paper, we consider a control-related service which is provided by a semi-honest cloud. This means that the cloud is interested in the data, but honestly executes the service. This assumption is reasonable because if the cloud behaves maliciously (i.e., deviates from the service protocol), the user would turn away from it. 
Now, in order to make the following analysis more general, our focus is (mostly) restricted to the input and output data of the cloud. The actual service (in the form of an algorithm) and the corresponding additional data is treated as a black-box, in spite of being potentially useful for an attack. 

More precisely, we denote the input plaintext data with $\xb$, which is encrypted according to $\yb=\Enc(\xb)$ and provided to the cloud. The former stem from the message space $\Mc^n$, whereas the latter come from the, typically larger, ciphertext space $\Cc^n$. The mapping between these spaces is the encryption $\Enc(\cdot)$, which will be specified soon. After receiving $\yb$, the cloud executes the service resulting in $\zb\in\Cc^m$ which is then sent to the user and decrypted according to $\ub=\Dec(\zb)\in\Mc^m$, where $\Dec(\cdot)$ reverses $\Enc(\cdot)$. Lastly, $\ub$ and $\xb$ can have some sort of relation. Here, we consider the special case when the cloud is interconnected with a linear (discrete-time) system
\begin{align}
\label{eq:dynamics}
    \xb(k+1)=\Ab\xb(k)+\Bb\ub(k),
\end{align}
where $\Ab\in\R^{n\times n}$ and $\Bb\in\R^{n\times m}$ are the system and input matrices, respectively. Among other examples, this case is practically relevant in linear model predictive control.
To distinguish between different queries to the cloud service that occur at different points in time, we identify each of them with a time step $k\in\N$ (including $0$) as in~\eqref{eq:dynamics}.

\subsection{Security definition}
Because any new information may pave the way towards more knowledge and since it is impossible to address every possible attack, it is desirable that a cipher is not leaking information at all (or only a negligible amount). Only in that case, one can be sure about security. Because control engineers are likely unfamiliar with security notions, let us briefly explain and formalize this idea next.

Suppose we collect many (scalar) ciphertexts $y\in\Cc$. Then, we could obtain an approximation of $P(Y=y)$, i.e., the probability that the random variable $Y$ takes on the particular ciphertext value $y$, by creating a normalized histogram from the observed data. Each of these observations is generated by the (usually probabilistic) encryption procedure $Y = \Enc(X)$ given a particular plaintext $x$, which we denote by
$$
    P(\Enc(x)=y):=P(Y=y | X=x).
$$
Using this relationship between plaintexts and ciphertexts, we can describe the probability of the ciphertext $y$ as
$$
    P(Y=y)=\sum_{x\in\Mc} P(\Enc(x)=y) P(X=x),
$$
where $P(X=x)$ is the a priori plaintext distribution.
Now, if the encryption procedure is such that any two plaintexts $x_1,x_2\in\Mc$ result in the same ciphertext with equal probability, i.e.,
\begin{equation}
\label{eq:ind}
    P(\Enc(x_1)=y)=P(\Enc(x_2)=y)
\end{equation}
the distribution of the ciphertext $y$ becomes independent of the plaintext distribution by means of
\begin{align}
    P(Y=y)&=P(\Enc(x)=y) \sum_{x\in\Mc} P(X=x) \nonumber\\
    \label{eq:perfectsecurity2}
    &=P(\Enc(x)=y).
\end{align}
Whenever~\eqref{eq:perfectsecurity2} holds for all $y\in\Cc$ and all $x\in\Mc$, the cryptosystem is called ``perfectly secure'' [see \citep[p.~32]{katz2007}]. Intuitively, a perfectly secure cipher does not reveal information about the plaintext because every ciphertext could be an encryption of every plaintext. Note that~\eqref{eq:ind} describes a uniform distribution. We will build on this observation later when we specify how randomness is introduced into the encryptions.

Next, we define the statistical distance by
\begin{equation}
    \label{eq:stat}
    D=\frac{1}{2}\sum_{y\in \Cc} | P( \Enc(x_1)=y)-P(\Enc(x_2)=y) |,
\end{equation}
where the value $D\in[0,1]$ measures the similarity of two distributions. Obviously, the key ingredient to derive~\eqref{eq:perfectsecurity2} is~\eqref{eq:ind}, which connects the notion of perfect security with~\eqref{eq:stat}. Namely, an encryption scheme $\Enc(x)$ is perfectly secure if and only if $D=0$ for every $x_1,x_2\in\Mc$, which is also called perfect indistinguishability of ciphertexts.
Moreover, using the statistical distance allows quantifying the security of a cipher by its deviation from $D=0$.
In the worst case ($D=1$), the ciphertext distributions are disjoint, which allows distinguishing all ciphertexts.
\begin{defn}
\label{def1}
The ciphertexts of an encryption scheme $\Enc(x)$ 
are called perfectly distinguishable, if $D=1$ for some $x_1,x_2\in\Mc$.
\end{defn}

Lastly, there exist relaxations of perfect indistinguishability in order to develop more practical cryptosystems. A notable one is as follows.
\begin{defn}
\label{def2}
The ciphertexts of an encryption scheme $\Enc(x)$
are statistically indistinguishable if $D\leq\negl(\kappa)$ for all $x_1,x_2\in\Mc$.
\end{defn}
Here, $\negl(\kappa)$ is a negligible function in the security parameter $\kappa$. The defining feature of such functions is that, by increasing $\kappa$, they approach $0$ (from above) more quickly than any inverse polynomial. 
A consequence of statistical indistinguishability is that there exist ciphertexts that reveal information about the plaintext, but their occurrence is statistically insignificant, which makes the success probability of an attack extremely small.
Both perfect and statistical indistinguishability are information theoretic properties. Hence, no amount of computation can increase the statistical distance (see \citet{cramer2015secure}). Further relaxations, in which computations do play a role, are not of interest here.

\subsection{Attack models}
Next, we specify a potential attacker's capabilities in terms of knowledge.
To this end, let us introduce one of Kerckhoffs' principles first~\citep[Section~1.2]{katz2007}.
It states that a cipher should be designed assuming that ``the enemy knows the system''. Typically, in the context of cryptography, this means full knowledge about how the cipher works, i.e., key generation, encryption, and decryption. In the context of encrypted control, it is natural, from our point of view, to extend this idea to knowledge of the cyberphysical system. On the other hand, security through obscurity (by trying to hide information) is widely discouraged in cryptography.
It is crucial to note that one cannot control how much background information is available to the attacker, and that the cipher's security should not be dependent on that.
In conclusion, Kerckhoffs' principle justifies using information that may be unavailable in a realistic scenario for most attackers. 
In order to depict realistic scenarios in Section~\ref{sec:detanalysis}, though,
we try to keep the amount of used information as small as possible.

Throughout this paper, we consider ciphertext analysis and known-plaintext as well as known-plant attacks without the necessity to consider stronger attacker models.
{If the attacker has access to (an arbitrary amount of) plaintext-ciphertext pairs $(\xb,\yb)$, we refer to the scenario as a known-plaintext attack.}
{We further} define a ``known-plant attack'' as a scenario, in which ciphertexts as well as the plant's structure and (if useful) further plant knowledge are given.

\section{Random affine transformation ciphers}
\label{sec:RT}
In this section, a deterministic as well as probabilistic variant of the RT cipher is introduced, which are analyzed in the remainder of this paper.

\subsection{Deterministic variant}
\label{subsec:det}
The deterministic variant of the RT cipher uses
\begin{equation}
    \label{eq:randomenc}
    \yb(k)=\Enc\left(\xb(k)\right)=\Rb\xb(k)+\rb
\end{equation}
as an encryption of the plaintext $\xb(k)$,
where $\Rb\in\R^{n\times n}$ and $\rb\in\R^{n}$ are randomly chosen once as
specified later.
We note that the encryption~\eqref{eq:randomenc} is a deterministic algorithm. Namely, for a fixed pair of keys $(\Rb,\rb)$, the plaintext $\xb(k)$ is always mapped to the same ciphertext $\yb(k)$.
In order to decrypt the ciphertext state $\yb(k)$, $\Rb$ must be invertible,
such that we obtain
\begin{equation}
    \label{eq:randomdec}
    \xb(k)=\Dec\left(\yb(k)\right)=\Rb^{-1}(\yb(k)-\rb).
\end{equation}
Because the same keys are used for encryption and decryption, the RT cipher is a symmetric cryptosystem.
There exist setups with no need to decrypt $\yb(k)$. For instance, $\yb(k)$ may provide an update for a quadratic program. Then, what a client receives is $\zb(k)$, which may be decrypted with different random keys $\Sb\in\R^{m\times m}$ and $\sb\in\R^m$ according to
\begin{equation}
    \label{eq:randomdec2}
    \ub(k)=\Dec\left(\zb(k)\right)=\Sb^{-1}(\zb(k)-\sb).
\end{equation}

\begin{rem}
We briefly remark the generalization in~\citet{hayati2022privacy}, where $\dim(\yb)>n$. This requires that the corresponding key $\tilde{\Rb}$ has full column-rank for decryption. However, here we restrict ourselves to quadratic $\Rb$, i.e., to less information. One can reduce this generalization to the quadratic case by selecting $n$ linearly independent values from $\yb$. In fact, the chance of linear dependence in the random matrix $\tilde{\Rb}$ is vanishingly small, and it would be detectable from $\yb$.
\end{rem}

\subsection{Probabilistic variants}
\label{subsec:prob}
An improvement of the security can be achieved by using non-constant keys $(\Rb(k),\rb(k))$ which are 
randomly sampled
in every time step.
To this end, we define two schemes. First, only $\rb(k)$ is non-constant. 
Applying this idea to the previous encryptions and decryptions yields
\begin{align}
\label{eq:r(k)}
    \yb(k)&=\Rb \xb(k)+\rb(k), \\
    \nonumber
    \xb(k)&=\Rb^{-1}\left(\yb(k)-\rb(k)\right) \\ 
\label{eq:q(k)}
    \ub(k)&=\Sb^{-1}\left(\zb(k)-\sb(k)\right).
\end{align}
Second, both $\Rb(k)$ and $\rb(k)$ are non-constant. Similarly to above, we obtain 
\begin{align}
\label{eq:R(k)}
    \yb(k)&=\Rb(k) \xb(k)+\rb(k), \\
    \nonumber
    \xb(k)&=\Rb(k)^{-1}\left(\yb(k)-\rb(k)\right) \\
\label{eq:Q(k)}
    \ub(k)&=\Sb(k)^{-1}\left(\zb(k)-\sb(k)\right).
\end{align}
Due to $\Rb(k)$, the latter results in
substantially more
communication and precomputation cost.
As opposed to before, these encryptions are probabilistic. For instance, $\xb(k)$ can be mapped to different $\yb(k)$ depending on the random vector $\rb(k)$ or both of the keys $(\Rb(k),\rb(k))$.

\section{Analysis of the deterministic variant}
\label{sec:detanalysis}
Before we dive into the analysis,
let use briefly note that we deliberately choose $n>1$ in this section, although the security of a cipher should also be guaranteed if only single bits were encrypted. 

\subsection{Ciphertext analysis}
\label{subsec:detcipheronly}
In many of the publications cited above, it is assumed that the deterministic RT operates under a ciphertexts-only setup. So, let us begin by considering the ciphertexts $\{\yb(k)\}_{k\in\N}$ arising from~\eqref{eq:randomenc}.
Equivalently, one could focus on~\eqref{eq:randomdec} or~\eqref{eq:randomdec2} for the analysis.
Restricting ourselves to only ciphertext information seems to make a concrete attack impossible, and the keys as well as $\xb$ remain ambiguous. This is because~\eqref{eq:randomenc} does not enable to check whether guessed keys are correct without any knowledge about plaintexts.
Still, an analysis of the ciphertexts by measuring the security through~\eqref{eq:stat} can be provided.

Since the RT cipher, as it has been considered in various publications, is defined over the reals, but~\eqref{eq:stat} is defined for discrete random variables, we follow~\citet{regev2009lattices} and consider an analogous expression for continuous variables, i.e.,
\begin{equation}
    \label{eq:statistical-distance-continuous}
    D=\frac{1}{2}\int_{\Cc^p} \big| p(\yb(k) \,|\, \xb_1(k)) - p(\yb(k) \,|\, \xb_2(k)) \big| \mathrm{d}\yb,
\end{equation}
where $p(\yb(k) \,|\, \xb(k))$ takes the role of $P(\Enc(x)=y)$ from before. This allows to state the following result.

\begin{prop}
Ciphertexts resulting from the encryption~\eqref{eq:randomenc} result in a statistical distance of $D=1$.
\end{prop}
\begin{proof}
As the encryption $\Rb\xb(k)+\rb$ of a given $\xb(k)$ always results in the same $\yb(k)$ as soon as the keys $(\Rb,\rb)$ are fixed, the corresponding conditional probability density takes the form of a Dirac delta impulse. With this at hand,~\eqref{eq:statistical-distance-continuous} reduces to
\begin{align*}
    D = \frac{1}{2}\int_{\Cc^p} \big| \delta(\yb(k) - \Rb\xb_1(k)) - \delta(\yb(k) - \Rb\xb_2(k)) \big| \mathrm{d}\yb=1
\end{align*}
for any given $\xb_1(k)\neq\xb_2(k)\in\Mc^n$. This lets us conclude that the RT cipher exhibits $D=1$ (perfect distinguishability according to Definition~\ref{def1}).
\end{proof}

This result is a consequence of the deterministic encryption and enables to break the cipher by means of known-plaintext attacks, as we will see in the next section.  
Interestingly,
it also forbids to make~\eqref{eq:randomenc}
a public key cryptosystem. To see this, note that with help of a public key, it would be possible to construct a large library of ciphertexts from chosen plaintexts. Then, because of the one-to-one relation of plaintexts and ciphertexts, these correspondences can be exploited by (approximately) matching them to observed ciphertexts.
Next, attacks on the deterministic variant \eqref{eq:randomenc}-\eqref{eq:randomdec2} are proposed, where we intend to infer more information than already available.

\subsection{Known-plaintext attack}
Under a known-plaintext attack, we have access to the plaintext and corresponding ciphertext sequences $\{\xb(k)\}_{k\in\N}$ and $\{\yb(k)\}_{k\in\N}$, respectively. Then, the encryption~\eqref{eq:randomenc} is linear in $(\Rb,\rb)$ with $n(n+1)$ unknowns. Now, for every time step, $n$ new equations are obtained. Therefore, with $(n+1)$ different plaintext-ciphertext pairs, one can uniquely solve for $(\Rb,\rb)$.
To this end, we plug the (not necessarily consecutive) sequences $\{\xb(k)\}_{k=0}^{n},\{\yb(k)\}_{k=0}^{n}$ into~\eqref{eq:randomenc} and rearrange into the form
\begin{equation}
    \label{eq:linearattack}
    \begin{pmatrix} 
    \begin{pmatrix} \xb(0)^\top  & 1 \end{pmatrix} \otimes \Ib_{n} \\ \vdots \\ \begin{pmatrix} \xb(n)^\top & 1 \end{pmatrix}\otimes \Ib_{n}  
    \end{pmatrix} 
    \!\!
    \begin{pmatrix} 
    \vect{\Rb}\\ \rb 
    \end{pmatrix}
    \!
    =
    \!
    \begin{pmatrix} 
    \yb(0) \\ \vdots \\ \yb(n)
    \end{pmatrix}\!
    =:\Upsilonb_{n+1},
\end{equation}
which is compatible with standard linear equation solvers.
Having extracted
the keys $(\Rb,\rb)$, it is possible to decrypt all ciphertexts $\yb(k)$.

As an alternative, if $p>n$ plaintext-ciphertext pairs are available, one might prefer building on the least squares approach
\begin{equation}
    \label{eq:leastsquares}
    \min_{\Rb,\rb}\sum_{k=0}^{p-1}\|\Rb \xb(k)+\rb-\yb(k)\|_2^2,
\end{equation}
where the optimal choice is given by
\begin{align*}
    \begin{pmatrix}\vect{\Rb} \\ \rb\end{pmatrix}=\Hb^+\Upsilonb_{p} 
    \text{, with }  
    \Hb:=\begin{pmatrix}
       \begin{pmatrix} \xb(0)^\top  & 1 \end{pmatrix} \otimes \Ib_{n} \\
       \vdots \\
       \begin{pmatrix} \xb(p-1)^\top  & 1 \end{pmatrix} \otimes \Ib_{n}
    \end{pmatrix}\!.
\end{align*}
The appeal of this variant is that possible errors or uncertainties in the plaintexts $\xb(k)$ tend to cancel out.
For instance, if $\yb(k)$ is an encryption of $\xb(k)$ plus Gaussian measurement noise, then~\eqref{eq:leastsquares} provides an optimal estimate which improves with additional data.

\subsection{Known-plant attack}
In the next step, the attacker's knowledge is reduced to structural knowledge about the system~\eqref{eq:dynamics} and the ciphertexts sequences $\{\yb(k)\}_{k\in\N}$ and $\{\zb(k)\}_{k\in\N}$. Here,~\eqref{eq:randomdec} is not used according to the setup (see Section~\ref{sec:setup}).
First, we get rid of the constant noise terms $\rb$ and $\sb$ by taking the difference between two ciphertexts, i.e., 
$$
\Delta\yb(k):=\yb(k+1)-\yb(k)=\Rb \Delta \xb(k)
$$
and likewise for $\Delta \zb(k)$. Then, combining all relations yields the ciphertext dynamics
$$
    \Delta\yb(k+1)=\Rb\Ab\Rb^{-1} \Delta \yb(k)+\Rb\Bb\Sb^{-1}\Delta\zb(k).
$$
Now, since the attacker has access to $\{\Delta\yb(k)\}_{k\in\N}$ and $\{\Delta\zb(k)\}_{k\in\N}$, it can identify $\Vb:=\Rb\Ab\Rb^{-1}$ as well as $\Wb:=\Rb\Bb\Sb^{-1}$ after collecting $n$ different ciphertext pairs by solving the linear system
\begin{align*}
    \begin{pmatrix}
    \Delta \yb(0)^\top \otimes \Ib_n  &  \Delta \zb(0)^\top \otimes \Ib_n\\
    \vdots & \vdots \\
    \Delta \yb(n-1)^\top \otimes \Ib_n & \Delta \zb(n-1)^\top \otimes \Ib_n
    \end{pmatrix}
    \begin{pmatrix} \!\!\!
    \vect{\Vb} \\ \vect{\Wb}
    \end{pmatrix}\!
    =\! \Upsilonb_n.
\end{align*}
Here, similarly to~\eqref{eq:linearattack} and~\eqref{eq:leastsquares}, a least-squares variant can be used whenever measurement noise is present.
Remarkably, due to
\begin{align*}
\det\left(\Vb-\lambda\Ib\right)&=\det\left(\Rb\Ab\Rb^{-1}-\Rb\lambda \Ib \Rb^{-1}\right)\\
&=\det\left(\Rb\right)\det\left(\Ab-\lambda \Ib\right)\det\left(\Rb^{-1}\right),
\end{align*}
$\Vb$ and $\Ab$ share the same characteristic polynomial (since they are related by a similarity transformation). Inevitably, this reveals the eigenvalues of $\Ab$.

Without further information, the plaintexts, keys, or plant matrices are ambiguous. An interesting case arises, though, if the attacker is able to reconstruct $\Ab$, e.g., by modeling the plant by means of $\Vb$. Then, due to the shared $n$ eigenvalues between $\Vb$ and $\Ab$, the Sylvester equation $\Vb\Rb-\Rb\Ab=\zerob$, rewritten in the form
$$
    \left(\Ib_n \otimes \Vb-\Ab^\top \otimes \Ib_n \right)\vect{\Rb}=\zerob,
$$
contains non-trivial solutions.
Although we cannot uniquely\footnote{Another Sylvester equation $\tilde{\Vb}=\Rb\tilde{\Ab}\Rb^{-1}$ and a scalar equation would be needed to identify $\Rb$ uniquely.} 
identify $\Rb$ from $\Vb$,
the null space characterized by the matrix
$$
\Nb=\mathrm{null}\left(\Ib_n \otimes \Vb-\Ab^\top \otimes \Ib_n \right)
$$
provides a lower dimensional search space for $\Nb \alphab=\vect{\Rb}$, where $\alphab\in\R^{n}$. 
Then, using $\Rb\Delta\xb(k)=\Delta\yb(k)$ one obtains
$$
\left(\Delta\xb(k)^\top \otimes \Ib_n\right)\Nb\alphab=\Delta\yb(k).
$$ 
This relation allows the reconstruction of $\Rb$ from $\alphab$ using only one correct plaintext-ciphertext pair, with $\Delta\xb(k)\neq\zerob$. 
From here on, recovery of $\rb$ is trivial, which allows reconstructing all $\xb(k)$ and $\Bb\ub(k)$.

\section{Analysis of the probabilistic variant}
\label{sec:probattack}
Next, we consider the probabilistic variants of the RT cipher.
To this end, we focus on~\eqref{eq:r(k)} and~\eqref{eq:R(k)} rather than~\eqref{eq:q(k)} and~\eqref{eq:Q(k)}. 
Again, the former relations provide the same structure as the latter and, therefore, yield the same security results.

\subsection{Ciphertext analysis I}
Now we are interested in whether information can be inferred from the ciphertexts $\{\yb(k)\}_{k\in\N}$ stemming from~\eqref{eq:r(k)}. As opposed to Section~\ref{subsec:detcipheronly}, the encryption $\Rb\xb(k)+\rb(k)$ can now output different ciphertexts for the same $\xb(k)$ and fixed $\Rb$ due to the time varying $\rb(k)$.
To quantify the security of this cipher, an expression for the integrand in~\eqref{eq:statistical-distance-continuous} is needed, which leads us to the sampling of elements in $\rb(k)$. By virtue of the observation made by~\eqref{eq:ind}, we propose a uniform distribution $p(\rb(k)) = \Uc(-r_{\max}\oneb, r_{\max}\oneb)$, which provides a probability of
\begingroup
\renewcommand*{\arraystretch}{1.3}
$$
\Uc(\xi; a , b) := \left\lbrace \begin{array}{cl}
          \frac{1}{b - a} &\quad \text{if}\; a \leq \xi \leq b, \\
          0 &\quad\text{otherwise}
      \end{array} \right.
$$
\endgroup
for every component of $\rb(k)$ parametrized by the finite upper bound $r_{\max}$. Then, one can state the following.

\begin{prop}
Let the components of $\rb(k)$ be independent and uniformly distributed in $[-r_{\max}, r_{\max}]$, for some finite $r_{\max}>0$,
and let $\xb_1(k),\xb_2(k)$ be arbitrary but fixed plaintexts inside $\Mc^n$.
Then, the $i$-th ciphertext component resulting from the encryptions~\eqref{eq:r(k)} provides a statistical distance of
\begin{equation}
\label{eq:statprob2}
    D=\frac{1}{2r_{\max}} \left|\left(\Rb\xb_1(k)-\Rb\xb_2(k)\right)_i\right|
\end{equation}
if $\left|\left(\Rb\xb_1(k)-\Rb\xb_2(k)\right)_i\right|\leq 2r_{\max}$ and $D=1$ otherwise.
\end{prop}
\begin{proof}
First, without any effect on the probability distribution $p(\yb(k)|\xb(k))$, we substitute $\tilde{\xb}(k):=\Rb\xb(k)$. Then, the encryptions are $\yb(k)=\tilde{\xb}(k)+\rb(k)$. Using $p(\rb(k)) = \Uc(-r_{\max}\oneb , r_{\max}\oneb)$ and~\eqref{eq:r(k)}, we see that $\yb(k)$ is distributed like $\rb(k)$, except for the mean being shifted by $\tilde{\xb}(k)$. Therefore, we obtain
\begin{equation*}
p(\yb(k) \,|\, \tilde{\xb}(k))
    = \Uc(-r_{\max}\oneb+\tilde{\xb}(k) , r_{\max}\oneb+\tilde{\xb}(k)).
\end{equation*}
Finally, the result follows from substituting
this probability into~\eqref{eq:statistical-distance-continuous} for $\tilde{\xb}_1(k)$ and $\tilde{\xb}_2(k)$, and evaluating the integral.
\end{proof}

The simple yet significant
substitution step reveals that $\Rb$ is not contributing to the security according to~\eqref{eq:stat} while it is used in~\eqref{eq:r(k)} to encrypt $\xb(k)$. Furthermore, we learn from~\eqref{eq:statprob2} that big $r_{\max}$ are crucial for the security of~\eqref{eq:r(k)}, where big values in $\Rb$ tend to have a counterproductive effect.
A remarkable consequence is, that~\eqref{eq:r(k)} can not be perfectly indistinguishable (perfectly secure) when finite $r_{\max}$ are used.
However, statistical indistinguishability is possible as follows.
First, one picks a finitely large sampling interval for the elements of $\Rb$ (or simply set $\Rb=\Ib$ if compatible). Then, the maximal noise is chosen as an exponential function in the security parameter, e.g., $r_{\max}=2^{\kappa}$. Evidently, the statistical distance becomes $D\propto 1/2^\kappa=\negl(\kappa)$ as required by Definition~\ref{def2}.
\begin{rem}
At this point, we note that~\eqref{eq:r(k)} (with $\Rb=\Ib$) can be thought of as an additive real number secret sharing scheme without modulo arithmetic. In fact, this is related to the results presented in~\citet{tjell2021privacy}, where a more general secret sharing scheme (Shamir's secret sharing) over real numbers is proposed. The differences between this and our work are as follows. Here, we build on definitions based on the statistical distance for security and on a uniform distribution over an interval. In~\citet{tjell2021privacy}, the authors quantify information leakage by the mutual information, which can be interpreted as the reduction of uncertainty about a plaintext given the corresponding ciphertext, and they use Gaussian noise distributions on $(-\infty,\infty)$. Interestingly, both choices for the noise distribution maximize the entropy for the respective case~\citep[p.412]{cover1999elements}.
\end{rem}

\subsection{Ciphertext analysis II}
For ease of presentation,
we limit ourselves to scalar quantities (i.e., $n=1$) in the remaining ciphertext analysis. 
In particular, we consider $y$, $x$, $R$, $r$, and $\tilde{x}$ as representatives for
$\yb$, $\xb$, $\Rb$, $\rb$, and $\Rb\xb$, respectively.
Next, we extend the analysis of the previous section by (a scalar version of)~\eqref{eq:R(k)} instead of~\eqref{eq:r(k)}.

\begin{prop}
Let $r(k)$ and $R(k)$ be independent and uniformly distributed on $[-r_{\max},r_{\max}]$ and $[-R_{\max},R_{\max}]$, respectively, with finite $r_{\max},R_{\max}>0$. Furthermore, $x_1(k),x_2(k)$ are arbitrary but fixed plaintexts inside $\Mc$ (excluding $0$) and fulfill $r_{\max}\geq |x_i(k)|R_{\max}$ for $i\in\{1,2\}$.
Then, scalar ciphertexts resulting from the encryption~\eqref{eq:R(k)} provide a statistical distance of
\begin{equation*}
    D
    = \frac{R_{\max}}{4 r_{\max}}\big||x_1(k)| - |x_2(k)|\big|.   
    \label{eq:stat-dist-R-resampled}
\end{equation*}
\end{prop}
\begin{proof}
By assumption, we have $p_r(r(k))=\Uc(-r_{\max},r_{\max})$ and $p_R(R(k))=\Uc(-R_{\max},R_{\max})$. Consider now
$\tilde{x}(k) = R(k)x(k)$,
which leads us to the uniform distribution
\begin{align*}
p_{\tilde{x}}(\tilde{x}(k)) 
=\Uc(-|x(k)|R_{\max}, |x(k)|R_{\max}).
\end{align*}
Then, by using the independence of $r(k)$ and $R(k)$, the conditional ciphertext distribution (with omitted dependence on $k$ for clarity) is
\begingroup
\renewcommand*{\arraystretch}{1.3}
\begin{align}
    \nonumber
    p(y \,|\, x)
    &= \int_{\Cc} p_{\tilde{x}}(y-\tau)  p_r(\tau)\,\mathrm{d}\tau \\
    &= \label{eq:conditional-ciphertext-dist-R(k)-scalar}
    \left\lbrace \begin{array}{cl}
        0                &\;\text{if } y < -(r_{\max} + |x| R_{\max}) \\
        f(y)             & \;\text{if } |y + r_{\max}| \leq |x| R_{\max} \\
        (2r_{\max})^{-1} & \;\text{if } |x| R_{\max} - r_{\max} < y \leq 0 \\
        p(-y \,|\, x)    & \;\text{if } y > 0,
    \end{array} \right.
\end{align}
\endgroup
with the abbreviation
$$
    f(y) := \frac{y(k) + r_{\max} + |x| R_{\max}}{4 r_{\max} R_{\max} |x|}.
$$
By substituting~\eqref{eq:conditional-ciphertext-dist-R(k)-scalar} into~\eqref{eq:statistical-distance-continuous} for $x_1(k)$ and $x_2(k)$ and evaluating the integral, we obtain the desired result.
\end{proof}

Here, the results from before are partially repeated, i.e., that big $r_{\max}$ provide security, whereas big $R_{\max}$ are counterproductive, and statistical indistinguishability is achievable by fixing $R_{\max}$ to a finite value and using an exponentially large $r_{\max}$.
The difference is now, however, that for the impractical case of $|x_1(k)| = |x_2(k)|$, a statistical distance of $D=0$ is attained. Moreover, the statistical distance is limited to the interval $D\in[0,{1}/{4})$ in this variant of the RT cipher.

Since both of these schemes allow, in principle, for parameter choices that make the chance of a successful attack vanishingly small, it is not expedient to propose attacks. However, the question arises, how the security carries over from these real number schemes to a more realistic floating point implementation. Because our analysis identified $\rb(k)$ as the security providing parameter, we investigate a scalar variant of~\eqref{eq:r(k)} in what follows.

\section{Floating point analysis}
\label{sec:floatsec}
\vspace*{-1mm}
Our previous analysis and the corresponding literature assume that $y, x, R, r$ are elements of $\R$. 
However, on a digital machine, there is only a finite amount of memory available and, thus, real numbers are not realizable.
Consequently, one has to settle with an approximation by using a finite set of numbers with finite precision. Before moving on, let us briefly note, that this simple observation allows concluding that any distribution based on such an approximation can not provide infinite entropy (as real numbers do), despite this is used as a basis for security arguments by some authors.

\subsection{Key generation} 
\label{subsec:keygen}
Next, we specify a (possible) key generation procedure and thereby fill the gap that is left by literature between a theoretical and practical implementation.
For a digital implementation, real numbers are most commonly emulated by floating point numbers (floats) based on the prevalent IEEE 754 standard.
It is well-known that a (normalized) float $x_f$ is then represented by 
\begin{equation*}
    \label{eq:float}
    x_f=(-1)^{b_{\mathrm{sign}}} \left(1+\sum_{i=1}^{M} b_{M-i} 2^{-i}\right)\times 2^{e},
\end{equation*}
where $b_{\mathrm{sign}}$, $b_{M-i}$, and $e$ are the sign bit, fraction bits, and exponent, respectively. Now, it seems likely that many authors have sampled from and computed with some set of floats
$$
    \F_{b}=\{x\in\R\,|\,\exists x_f: x-x_f=0\},
$$
where $b$ stands for a particular choice for {the number of bits that are occupied in total by $b_{\mathrm{sign}}$, $b_{M-i}$, and $e$ in the float representation of $x_f$}. By proceeding in the same manner, the message and ciphertext space, $\Mc$ and $\Cc$,
{actually consist of}
elements of $\F_{b}$.

\begin{figure}[tp!]
	\centering
	\includegraphics[trim=3.5cm 12.2cm 3.8cm 12.2cm, clip=true,width=.98\linewidth]{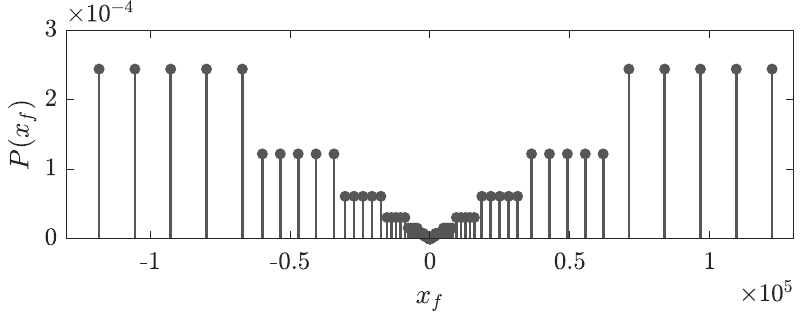}
    \vspace*{-2mm}
	\caption{Sampling probability of every $200$-th float from $\F_{16}$.
	}
	\label{fig:graphic1}
    \vspace{-5mm}
\end{figure}

Again, we want to sample from a uniform distribution. Remarkably, sampling elements from $\F_{b}$ uniformly at random will (with overwhelming probability) lead to very small numbers.
This can easily cause numerical problems, and it leads to undesired ciphertext distributions. The reason for that is that the interval $[2^{e},2^{e+1})$ doubles in size for every increment of $e$, while the amount of representable numbers
{is $2^{M}$ in every interval}.
Consequently, the elements in $\F_{b}$ (regardless of $b$) are very dense around $0$ and thin out exponentially as one moves away from $0$.
To counteract this effect, we propose (for the sake of analysis) a sampling probability $P(x_f)$ which doubles for every increment of $e$. To this end, note that $\alpha:=2^{1-B}$ is the smallest representable number ($2^{-B}$ is typically reserved), where $B$ is the exponent bias.
Hence, the number of increments in $e$ is
$$
\floor{\log_2(|x_f|)-\log_2(\alpha)}={\floor{\log_{2}(|x_f| \alpha^{-1})}},
$$
which is whenever $P(x_f)$ needs to double. Then, we find the desired float sampling probability 
$$ 
    P(x_f)
    =\gamma 2^{\floor{\log_{2}(|x_f| \alpha^{-1})}},
$$
which is illustrated in Figure~\ref{fig:graphic1}.
The factor $\gamma$
{ensures that}
$\sum_{x_f\in\F_b} P(x_f)=1$.
With this at hand, the resulting sampling distribution emulates the uniform distribution $\Uc(-\beta,\beta)$ over $\R$,  where $\beta:=(2-2^{-M})\times 2^B$ is the biggest number in $\F_b$.\footnote{Alternatively, one could use a standard pseudorandom function, that usually outputs numbers within $(0,1)$, and scale the result. However, in comparison to what we describe here, this will lead to a bigger spacing between the floats, which depends on the scaling.}

Since computations, performed by the cloud, are treated as a black-box here, we neglect numerical issues that can arise during computations with floats and other subtleties. For instance, special care needs to be taken such that overflows during the encryption and accuracy issues do not impair the result.

\subsection{Ciphertext analysis}
Finally, we turn our focus to the security of a scalar variant of the encryption~\eqref{eq:r(k)}, where ciphertexts $y(k)=\tilde{x}(k)+r(k)$ are computed with floats $y(k),\tilde{x}(k),$ and $r(k)$.
Due to the discrete nature of floats, the definition~\eqref{eq:stat} for the statistical distance is used here, which yields
\begin{equation}
    \label{eq:floatdist}
    D=\frac{1}{2}\sum_{y\in \F_b} | P(y(k)|\tilde{x}_1(k))-P(y(k)|\tilde{x}_2(k)) |.
\end{equation}
Similar to before, the encryption of $\tilde{x}(k)$ leads to $y(k)$ only if $Q_{\F_b}(\tilde{x}_i(k)+r(k))=y(k)$, where $Q_{\F_b}$ denotes the rounding procedure defined for $\F_b$. As a result, there may be multiple $r\in[-r_{\max},r_{\max}]\cap\F_b$ that can lead to the same $y(k)$ {for a given $\tilde{x}_i(k)$} or no such $r$ exists. Correspondingly, the (possibly empty) set $\Rc$ collects all
{such $r$ for a particular $y(k)$.}
With this at hand, we can now state the probability $P(y(k)|\tilde{x}_i(k))$, which is the probability that any of the $r\in\Rc$ is selected. More precisely,
\begin{equation}
    \label{eq:floatprob}
    P(y(k)|\tilde{x}_i(k))=
    \sum_{r\in\Rc} \gamma 2^{\floor{\log_{2}(|r| \alpha^{-1})}}.
\end{equation}

\textit{Example.} Due to the nonlinearities in $P(y(k)|\tilde{x}_i(k))$ an analytical expression for $D$ seems to be out of reach. Still, we are interested to see if the security results {for real numbers} from before carry over to a float based implementation, which motivates the following numerical study.
To get a handle on the cardinality of float sets, we opted for the half-precision format which we denote by $\F_{16}$, where $M=10$ and $e$ is represented with $5$ bits.
This allows to evaluate~\eqref{eq:floatdist} for all $y\in{\F_{16}}$, where we use~\eqref{eq:floatprob} for different choices of $\tilde{x}_1(k)$, $\tilde{x}_2(k)$, and $r_{\max}$.

\begin{figure}[tp!]
    \centering
    \includegraphics[trim=3.5cm 9cm 3.7cm 10cm, clip=true,width=.98\linewidth]{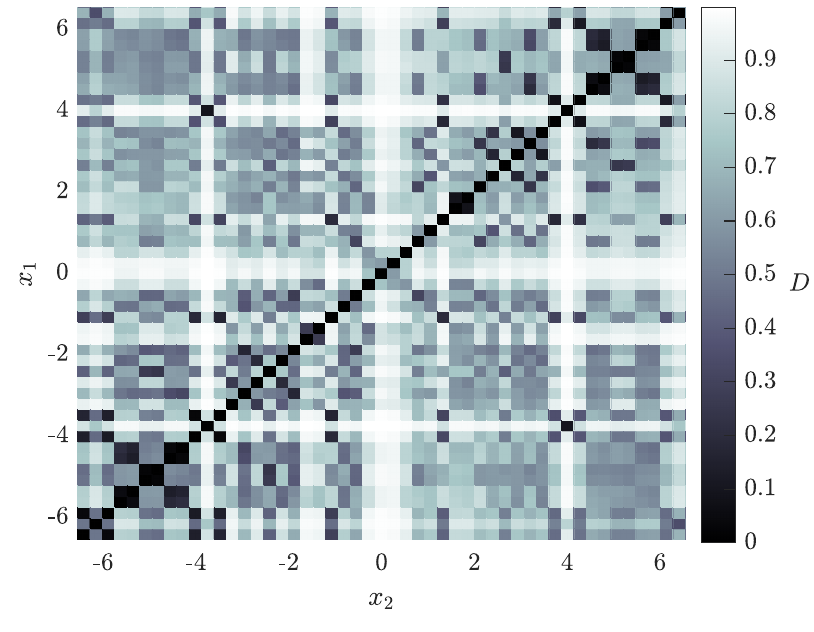}
    \vspace*{-4mm}
    \caption{Statistical distance of $100$ linearly spaced values for $\tilde{x}_1(k),\tilde{x}_2(k)\in[-0.01\% \beta,0.01\% \, \beta]\cap\F_{16}$ and all $y(k)\in\F_{16}$, where $r_{\max}=Q_{\F_{16}}\left(99.99\% \, \beta\right)$.}
    \label{fig:statisticaldistance}
    \vspace{-5mm}
\end{figure}

In order to see if a secure (and somewhat practical) implementation is achievable, we set $r_{\max}=Q_{\F_{16}}\left(99.99\% \, \beta\right)$, i.e., $99.99\%$ of the ciphertext space is used for noise and $0.01\%$ is left for the message space $\Mc$. As a result, $\tilde{x}_1$ and $\tilde{x}_2$ are roughly elements of $[-6,6]\cap\F_{16}$. The resulting statistical distances are depicted in Figure~\ref{fig:statisticaldistance}. Note
that the statistical distances start at $D=1$ for $r_{\max}=0$ and, by increasing $r_{\max}$ until $Q_{\F_{16}}\left(99.99\% \, \beta\right)$, evolve into Figure~\ref{fig:statisticaldistance} without changing shape.
As with a real number scheme, an increase of $r_{\max}$ tends to lead to smaller statistical distances. However, the connection {seems to be} much more complex than in~\eqref{eq:statprob2} which {makes} the generation of a secure instance of~\eqref{eq:r(k)} a very intricate task.
This is especially problematic since very low statistical distances can be right next to very high statistical distances in the message space. 
Apart from this overall diagnosis, let us highlight two things. First, as expected, $D=0$ along the diagonal (where $\tilde{x}_1(k)=\tilde{x}_2(k)$).
Second, different from the real number scheme, the maximum statistical distance occurs when either $\tilde{x}_1(k)\approx0$ or $\tilde{x}_2(k)\approx0$ and they have some distance between them (the ``white cross'' in Figure~\ref{fig:statisticaldistance}). Interestingly, this effect stems from the finite precision of float numbers. Suppose $\tilde{x}_1(k)>\tilde{x}_2(k)\approx0$, then, in order to obtain the same ciphertext $y(k)$, there must exist $r_i\in[-r_{\max},r_{\max}]\cap\F_b$ for $i\in\{1,2\}$ such that $Q_{\F_{16}}\left(\tilde{x}_1(k)+r_1\right)=Q_{\F_{16}}\left(\tilde{x}_2(k)+r_2\right)$. However, if $\tilde{x}_1(k)$ is large, the only possibility to obtain a small ciphertext is to use a large negative $r_1$. In that case, it is however impossible to output every possible $y\in\F_{16}$ around zero because $r_1$ does not provide enough precision as a result of its size. Hence, small ciphertexts can only be a result from relatively small plaintexts. Of course, these effects lose severity when computations are carried out over single-precision numbers from $\F_{32}$ or double-precision numbers from $\F_{64}$ but they are still present. 

\section{Conclusion}
\label{sec:outlook}
\vspace*{-1mm}
In this paper, we provide a cryptanalysis for random affine transformation ciphers that recently gained attention especially for solving control-related optimization problems as in model predictive control. To this end, a generalized setup compatible with many schemes from the literature, as well as security definitions and different attack models are introduced. Using this framework, two variants of the random affine transformation cipher are analyzed. 
First, the deterministic variant, which reuses the encryption keys, is shown to be not secure according to its statistical distance. As a result, the cipher breaks completely under a known-plaintext attack and leaks useful information about the closed-loop in which it is used without plaintext knowledge. Second, the achievable security of the probabilistic variant was a surprise to us. Opposed to the presentation in the paper, we first attacked the scheme (with partial success) and painfully realized that this was only possible because elements of $\Rb$ and $\rb$ were sampled from similar intervals. A subsequent ciphertext analysis revealed that the real number scheme can achieve statistical indistinguishability if the keys are selected according to our results. For a more practical insight, we translated the scheme to floating point numbers by defining a key generation procedure. Then, the numerical security analysis reveals a much more complex security relation which hinders a practical application, where security guarantees are needed. 

Finally, we conclude by comparison with its real number variant, that possible attacks against a floating point implementation of the probabilistic random affine transformation cipher should exploit the floating point peculiarities. For instance, rounding makes certain ciphertexts impossible or unlikely and, according to our floating point sampling distribution, the floating point exponent distribution is non-uniform. Moreover, using the scheme inside a control-loop leads to ``encrypted dynamics'', where the uniformly distributed noise approximately becomes Gaussian if the information is aggregated over several time steps.

\section*{Acknowledgment}
Financial support by the German Research Foundation (DFG) and the Daimler and Benz Foundation under the grants SCHU 2940/4-1 and 32-08/19 is gratefully acknowledged.

\bibliographystyle{ifacconf}

\end{document}